\RequirePackage{lineno}
\documentclass[aps,prl,10pt,twocolumn,superscriptaddress,longbibliography]{revtex4-2}

\usepackage{graphicx}
\usepackage{amssymb,amsmath,amsfonts,gensymb}
\usepackage{bm,hyperref}
\usepackage{color}
\usepackage{ulem}
\usepackage{bbm}
\usepackage{subfigure}
\usepackage{dcolumn}
\usepackage{mathtools}
\usepackage{pifont}

\def\I{\uppercase\expandafter{\romannumeral 1}}
\def\II{\uppercase\expandafter{\romannumeral 2}}
\def\III{{\uppercase\expandafter{\romannumeral 3}}}
\def\IV{{\uppercase\expandafter{\romannumeral 4}}}
\def\V{{\uppercase\expandafter{\romannumeral 5}}}
\def\VI{{\uppercase\expandafter{\romannumeral 6}}}
\def\VII{{\uppercase\expandafter{\romannumeral 7}}}
\def\i{\lowercase\expandafter{\romannumeral 1}}
\def\ii{\lowercase\expandafter{\romannumeral 2}}
\def\iii{{\lowercase\expandafter{\romannumeral 3}}}
\def\iv{{\lowercase\expandafter{\romannumeral 4}}}
\def\v{{\lowercase\expandafter{\romannumeral 5}}}
\def\vi{{\lowercase\expandafter{\romannumeral 6}}}
\def\vii{{\lowercase\expandafter{\romannumeral 7}}}

\newcommand{\eV}{{\text{eV}}}
\newcommand{\eVA}{{\text{eV}\cdot \text{\AA}}}
\newcommand{\bk}{\mathbf{k}}
\newcommand{\btk}{\widetilde{\mathbf{k}}}
\newcommand{\btq}{\widetilde{\mathbf{q}}}
\newcommand{\br}{\mathbf{r}}

\def\k{\mathbf{k}}

\def\G{\mathbf{G}}

\def\q{\mathbf{q}}
\def\Q{\mathbf{Q}}

\newcommand{\bG}{\mathbf{G}}
\newcommand{\bQ}{\mathbf{Q}}

\begin{document}

\title{General Many-Body Perturbation Framework for Moir\'e Systems}

\author{Xin Lu}
\affiliation{School of Physical Science and Technology, ShanghaiTech Laboratory for Topological Physics, State Key Laboratory of Quantum Functional Materials, ShanghaiTech University, Shanghai 201210, China}

\author{Yuanfan Yang}
\affiliation{School of Physical Science and Technology, ShanghaiTech Laboratory for Topological Physics, State Key Laboratory of Quantum Functional Materials, ShanghaiTech University, Shanghai 201210, China}

\author{Zhongqing Guo}
\affiliation{School of Physical Science and Technology, ShanghaiTech Laboratory for Topological Physics, State Key Laboratory of Quantum Functional Materials, ShanghaiTech University, Shanghai 201210, China}

\author{Jianpeng Liu}
\email{liujp@shanghaitech.edu.cn}
\affiliation{School of Physical Science and Technology, ShanghaiTech Laboratory for Topological Physics, State Key Laboratory of Quantum Functional Materials, ShanghaiTech University, Shanghai 201210, China}
\affiliation{Liaoning Academy of Materials, Shenyang 110167, China}
	
\bibliographystyle{apsrev4-2}

\begin{abstract} 
    Moir\'e superlattices host a rich variety of correlated topological states, including interaction-driven integer and fractional Chern insulators. A common approach to study interacting ground states at integer fillings is the Hartree-Fock mean-field method. However, this method neglects dynamical correlations, which often leads to an overestimation of spontaneous symmetry breaking and fails to provide quantitative descriptions of single-particle excitations. This work introduces a general many-body perturbation framework for moir\'e systems, combining all-band Hartree-Fock calculations with $GW$ quasiparticle corrections and random phase approximation (RPA) correlation energies. We apply this framework to hexagonal boron nitride aligned rhombohedral pentalayer graphene and magic-angle twisted bilayer graphene (MATBG). We show that incorporating RPA correlation energy and $GW$ self-energy corrections yields phase diagrams and single-particle spectra that quantitatively align with experimental measurements for both systems. Particularly, the ground state at charge neutrality of MATBG is predicted to be a nematic metal, which is stabilized over Kramers intervalley coherent insulator due to lower correlation energy. Our versatile framework provides a systematic beyond-mean-field approach applicable to generic moir\'e systems.
    \end{abstract} 
    
    \maketitle

    \paragraph{Introduction}
    The advent of moir\'e superlattices represents a conceptual breakthrough in condensed matter physics: a small twist between two otherwise weakly correlated materials, such as graphene and transition metal dichalcogenides, can give rise to flat bands dominated entirely by $e$-$e$ interactions. Consequently, strongly correlated phenomena, including unconventional superconductivity \cite{cao-nature18-supercond} and correlated insulators \cite{cao-nature18-mott}, emerge in moir\'e platforms. Most strikingly, recent experiments have provided evidence for the fractional quantum anomalous Hall effects in twisted transition metal dichalcogenides \cite{fqah-nature23,fqah-prx23,fqah-optics-xu-nature23,fqah-mak-nature23} and in hexagonal boron nitride (hBN)-aligned rhombohedral $n$-layer graphene (R$n$G) \cite{fqah-julong-nature24,fqah-luxb-natmat25,fqah-julong-nature25}. This exotic many-body quantum state, termed  fractional Chern insulator (FCI) \cite{fci-prx11,sheng-fci-nc11,murdy-fci-prl11,wen-kagome-prl11,sarma-flatchern-prl11,qi-fqah-prl11}, constitutes a lattice realization of the fractional quantum Hall states of Landau levels and highlights the intricate interplay between topology and strong correlations.

    These remarkable discoveries have fostered the view that correlation effects are always essential in moir\'e systems. It is therefore striking that Hartree-Fock (HF) mean-field theory, which neglects dynamical fluctuations, often agrees well with experiments, especially at partial integer fillings. In magic-angle twisted bilayer graphene (MATBG), for example, HF has explained cascade transitions through non-rigid single-particle spectra \cite{Yazdani-cascade-nature20,Ilani-cascade-nature20,kang-cascade-tbg-prl21} and predicted isospin-polarized correlated insulators \cite{kang-tbg-prl19,xie-tbg-prl20,zaletel-tbg-hf-prx20,shang-nematic-prr21,Liu-magnetic-vp-tdbg-nc22,zhang-tbmg-prl22,Li-mao-tbmg-exp-nc22,liu-1storderTDBG-prx-2023,jpliu-tbghf-prb21,zhang-liu-tbg-prl22,efetov-nature19,young-tbg-science19,sharpe-science-19,efetov-nature20}. The correlation effects omitted in HF are those responsible for dynamical screening, which weakens the bare Coulomb interaction. Consequently, HF is typically biased to symmetry-breaking states, which generally fails to reproduce experimental measurements quantitatively. Methods such as constrained random phase approximation (RPA) \cite{pollet-tbg-crpa-prb20,wehling-tbg-prb19,zhang-liu-tbg-prl22,truncated-apw-prb23,fqah-luxb-natmat25} partially include screening, but typically only in a static way, missing its dynamical and full momentum-dependent character. Other beyond-mean-field techniques including density matrix renormalization group \cite{zaletel-dmrg-prb20,bultinck-tbg-strain-prl21,kang-tbg-dmrg-prb20}, exact diagonalization \cite{regnault-tbg-ed,macdonald-tbg-ed-prl21,yu-R5GED-prb-2025}, and quantum Monte Carlo \cite{hofmann-tbg_qmc-prx-2022,meng-tbg-qmc-cpl21,pan-meng-tbg-prl23,yuan-tbg_qmc-prx-2021,huang-tbg_qmc-natcomm-2025} are introduced to study correlated states in moir\'e systems. However, these methods face major limitations, including severe finite-size constraints when multiple bands are involved and sign problems for generic models or fillings. Analytical insights are available only for high-symmetry moir\'e models in certain ideal limits, such as MATBG in the ``chiral limit'' \cite{origin-magic-angle-prl19} and neglecting kinetic energy \cite{zaletel-tbg-hf-prx20,Bernevig-tbg3-arxiv20,lian-tbg-iv-prb21}, where the strong-coupling regime maps onto a quantum Hall ferromagnetism problem, but cannot be generalized to generic moir\'e systems.
    
    \begin{figure*}
        \centering
        \includegraphics[width=0.7\textwidth]{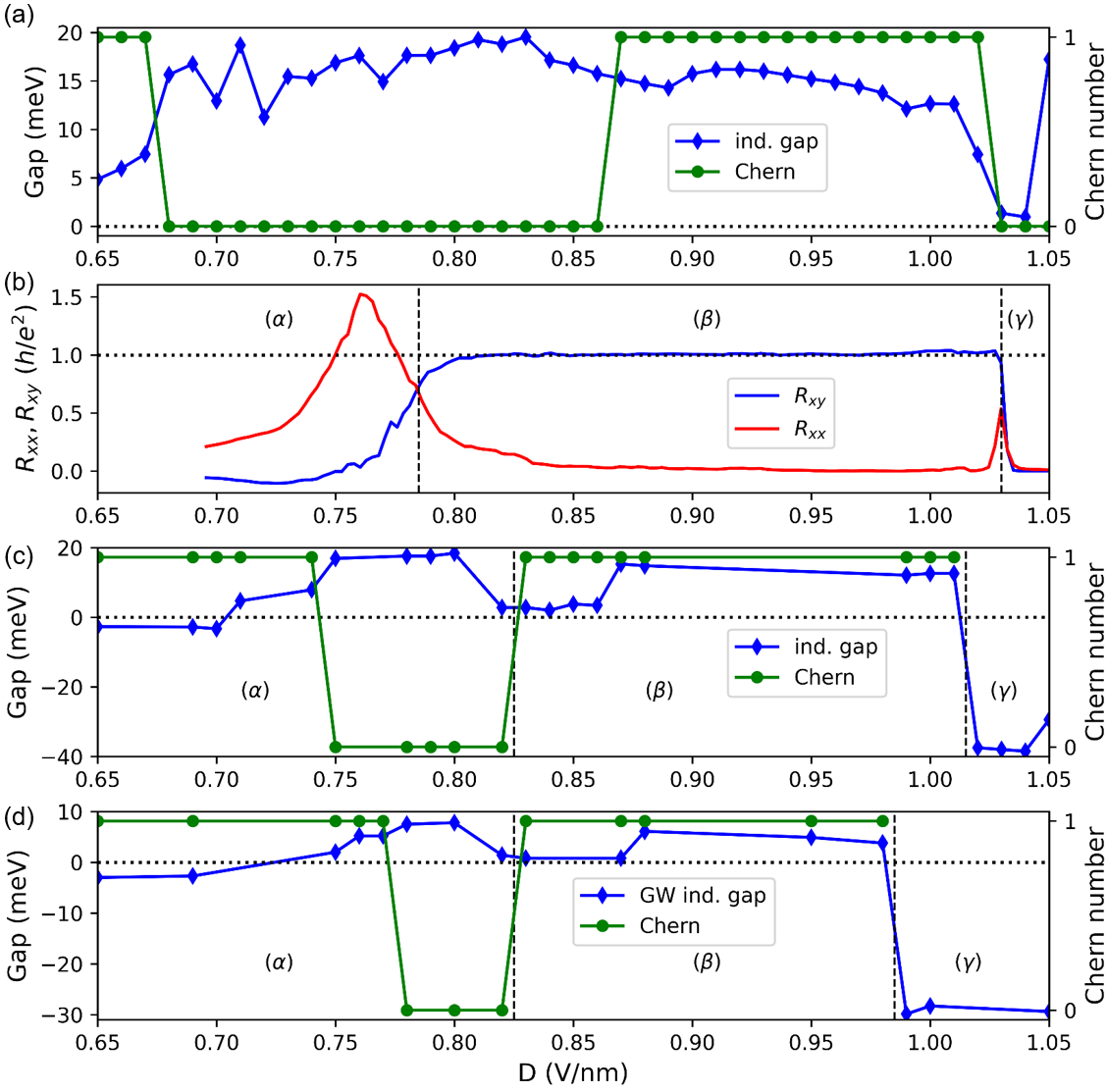}
        \caption{Comparison between (a) the evolution of the Chern number of the first conduction band and the indirect gap as a function of the $D$ field from the bare all-band HF calculations using $\epsilon_r=8$, (b) the evolution of transverse and longitudinal resistance observed experimentally for R5G-hBN \cite{fqah-julong-nature24,data-lu-fqah-nature-2024} and those obtained from (c) HF+RPA and (d) HF+$GW$+RPA calculations using the same $\epsilon_r=8$. Different phases are separated by vertical dashed lines \cite{data}. In the experimental phase diagram, the boundary between regions~$\alpha$ and $\beta$ is defined by $D$ at which the longitudinal resistance rises to half of its peak value.} 
        \label{fig:HFGWRPA}
    \end{figure*}
    
    In contrast, many-body perturbation theory offers a systematic framework in which correlation effects are treated perturbatively with respect to certain mean-field ground states. Most saliently, it can handle system sizes comparable to HF calculations and is applicable to generic moir\'e systems. The validity of perturbation theory requires only the absence of a phase transition upon inclusion of the dynamical correlation effects as a perturbation. Thus, provided that HF states qualitatively capture the experimentally observed phenomena, perturbation theory can enhance the quantitative accuracy of theoretical predictions in both ground-state energetics and single-particle excitation spectra. Moreover, it provides a means to assess the strength of correlation effects through the quasiparticle weight extracted from the single-particle self-energy, which measures the proximity of the many-body ground state to a Slater determinant.
    
    In this work, we introduce a general many-body perturbation framework that applies to generic symmetry-breaking states emerging in moir\'e systems. As a first step, we perform self-consistent HF calculations in the original plane-wave basis. In particular, our approach retains all moir\'e bands up to the plane-wave cutoff (typically $>1$\;eV) in the continuum model, together with the full spin and valley degrees of freedom. We refer to this scheme as the all-band HF method, in contrast to the band-projected HF approximations commonly used in the studies of moir\'e materials. By augmenting all-band HF with RPA correlation energies, we compare the RPA-corrected total energies of different HF-converged states to identify the many-body ground state at the RPA level. This is different from previous weak-coupling analyses in MATBG, where the RPA correlation energy was computed based on Hartree bands \cite{zhu-weakcoupleTBG-prb-2024}. Furthermore, we incorporate dynamical and inhomogeneous screening effects through the $GW$ approximation to the single-particle self-energies of generic symmetry-breaking states, which yields renormalized $GW$ single-particle spectra, that more faithfully capture experimentally measurable quantities such as energy gaps, Fermi velocities and bandwidths, which are not discussed in previous weak-coupling analyses, as far as we know. The RPA correlation energy can be further calculated based on $GW$ quasiparticle bands, which would give a more accurate description to the balance between exchange and correlation effects \cite{guo-AHC-arxiv-2025}. To demonstrate the power of our theoretical framework, we consider two typical moir\'e systems: hBN-aligned R5G and MATBG. For the hBN-aligned R5G system, our all-band HF calculations combined with RPA correlation energy and $GW$ self-energy corrections yield a phase diagram that is quantitatively consistent with the experimental one. For MATBG, we find that the ground state at charge neutrality is a nematic semimetal \cite{shang-nematic-prr21,bultinck-tbg-strain-prl21}, which is favored over Kramers intervalley coherent (K-IVC) state \cite{zaletel-tbg-hf-prx20} due to dynamical correlation effects. Remarkably, the calculated $GW$ quasiparticle bands around $\nu=0$ are quantitatively consistent with the results from quantum twisting microscopy \cite{xiao-TBG_QTM-arxiv-2025}.

    We first show the results for R5G-hBN heterostructures with twist angle $0.77^{\circ}$ with the configuration mimicking the experimental one \cite{fqah-julong-nature24}. The application of a displacement field $D$ pointing towards the moir\'e interface drives conduction band electrons towards the moir\'e-distant side. The non-interacting Hamiltonian uses Slater-Koster parameters \cite{koshino-prx18,moon-hbn-graphene-prb14,guo-HFfci-prb-2024,supp}, and the mapping between the vertical displacement field and the interlayer potential drop is determined self-consistently by iteratively solving for the electron distribution across the graphene layers \cite{guo-HFfci-prb-2024,supp}. While much theoretical attention has been given to the emergence of Chern insulator under strong $D$ using band-projected HF approaches \cite{dong-R5GAHC-prl-2024,dong-R5Gfci-prl-2024,zhou-R5Gfci-prl-2024,guo-HFfci-prb-2024,kwan-RnGHF-prb-2025}, few studies have discussed the full phase diagram at moir\'e filling $\nu=1$ across experimentally accessible displacement fields. This remains a fundamental yet overlooked problem, particularly since HF phase diagrams from different groups often do not align with each other due to the different treatments of Coulomb interaction normal ordering under band-truncated HF schemes. Our all-band HF scheme completely resolves this concern. In our calculations, we consider only the dominant intravalley Coulomb interactions. We use a dielectric constant $\epsilon_r$, which is the only fitting parameter in our theory, to account for all the static homogeneous screening effects.

    A comparison between the all-band HF phase diagram at filling $\nu=1$ for $\epsilon_r=8$ [Fig.~\ref{fig:HFGWRPA}(a)] with the experimental measurements [Fig.~\ref{fig:HFGWRPA}(b)] \cite{fqah-julong-nature24} reveals clear qualitative discrepancies. Experimentally, the device undergoes a sequence of phase transitions as the displacement field increases from $D=0.7$\;V/nm and $D=1.1$\;V/nm. Upon increasing $D$, the system first evolves from a metallic state to an insulating-like state. The longitudinal resistance exhibits a peak at $D=0.76$\;V/nm with a half-height width of approximately $0.04$\;V/nm, defining region~$\alpha$ in Fig.~\ref{fig:HFGWRPA}(b). As $D$ increases further, the system enters a Chern-insulator phase (region~$\beta$), characterized by quantized Hall resistance starting at $D \approx 0.8$\;V/nm. Finally, at $D=1.03$\;V/nm, the Hall resistance abruptly drops to zero, indicating a first-order transition back to a time-reversal symmetric metal, referred to as region~$\gamma$. In comparison, although the all-band HF results for $\epsilon_r=8$ exhibit a similar transition from a trivial insulator to a Chern insulator as $D$ increases from 0.68\;V/nm to 1.02\;V/nm [Fig.~\ref{fig:HFGWRPA}(a)], the bare HF phase diagram incorrectly yields a Chern insulator below 0.68\;V/nm and a trivial insulator above $D=1.02$\;V/nm. We emphasize that this discrepancy with experiment cannot be resolved, even qualitatively, by simply changing $\epsilon_r$, as confirmed by additional calculations using different $\epsilon_r$ \cite{supp}. It instead points to the essential role of dynamical screening and correlation effects beyond the HF mean-field description.

    To incorporate dynamical screening and correlation effects, we include in the total energy the RPA correlation energy, which captures the contribution of plasmonic collective excitations \cite{supp,fetter-book}. The negative RPA correlation energy $E_{c}^{\text{RPA}}$ favors states with stronger charge fluctuations, thus preferentially stabilizes metallic (or small-gap) states over gapped (or large-gap) states. By counteracting the exchange-driven tendency to favor gapped states, this correction yields a substantial improvement over the bare HF results. Fig.~\ref{fig:HFGWRPA}(c) presents the Chern number and the global gap as functions of displacement field, obtained by including $E_{c}^{\text{RPA}}$ in the ground-state energy evaluation of the HF-converged solutions for $\epsilon_r=8$. The resulting phase diagram can be divided into three regions, also labeled by $\alpha$, $\beta$ and $\gamma$, in close agreement with the experimental observations through comparing Fig.~\ref{fig:HFGWRPA}(b) and (c). First, region~$\gamma$ is identified as a time-reversal symmetric metal in the HF+RPA calculations, consistent with the experimentally observed abrupt drop of the anomalous Hall resistance  when $D\gtrapprox 1.03$\;V/nm. Remarkably, the theoretical transition occurs at $D=1.02$\;V/nm, demonstrating quantitative agreement with experiment. Second, the calculations predict a transition from a $C=1$ Chern insulator (region~$\beta$) to a trivial insulator when $D<0.82$\;V/nm. This is consistent with the experimental observation that the Hall resistance begins to deviate from the resistance quantum and the longitudinal resistance increases when $D$ decreases below approximately $0.8$\;V/nm. In addition, the predicted trivial insulator region is centered only $0.04$\;V/nm away from the experimentally observed longitudinal-resistance peak at $0.76$\;V/nm.

    The smooth evolution of the experimental resistance for $D<0.8$\;V/nm (region~$\alpha$) may arise from strong competition among several phases with similar energies, including a trivial insulator, a time-reversal breaking metal (with a $C=1$ first conduction band and negative indirect gap), a small-gap Chern insulator, and a time-reversal symmetric metal \cite{supp}. In realistic experimental conditions, extrinsic effects such as disorder and spatial inhomogeneity are expected to smear the phase boundaries, leading to crossovers rather than sharp transitions in transport measurements. Nevertheless, the overall experimental trend is well captured by our theoretical results: as $D$ decreases from $0.8$\;V/nm, the system evolves from a Chern insulator to a trivial insulator and eventually to a time-reversal breaking metallic state with small but finite Hall and longitudinal resistances. Overall, the HF+RPA calculations achieve quantitative agreement with experiment.

    We can recalibrate the HF+RPA phase diagram into an HF+$GW$+RPA phase diagram [Fig.~\ref{fig:HFGWRPA}(d)] by computing RPA correlation energy using $GW$ quasiparticle bands of different HF states under $\epsilon_r=8$. In the $GW$ approximation \cite{hedin-gw-pr65,louie-ppa-prb86,aryasetiawan-gw-rpp98,rubio-rmp02,reining-gw-wires18,golze-gw-fc19}, we replace the bare interaction $V$ in the HF self-energy with an RPA screened, frequency dependent interaction $W$, which captures the effects of couplings between single electron and collective charge fluctuations. In practice, we use eigenvalue-only $GW$ scheme \cite{surh-EVGW-prb-1991}. Calculations show that different $GW$ schemes lead to similar results \cite{supp}. Compared with the HF+RPA results, the HF+$GW$+RPA phase diagram retains the overall structure of the ground-state phases, with only minor quantitative differences. One notable improvement introduced by the $GW$ correction is the narrowing of the trivial-insulator region, yielding a reduced width of $0.05\;$V/nm, quantitatively consistent with the experimentally observed full width at half maximum of $0.04$\;V/nm for the longitudinal-resistance peak. In addition, the single-particle gaps and bandwidths are substantially reduced relative to the HF results. To illustrate the impact of the $GW$ correction on the quasiparticle spectra, Fig.~\ref{fig:R5G_HF_GW_eps8} shows the single-particle spectra of the ground states found at $D=0.65$, 0.80, 0.95 and 1.00\;V/nm with $\epsilon_r=8$, corresponding respectively to a time-reversal breaking metal and a trivial insulator in region~$\alpha$, a Chern insulator in region~$\beta$ and a time-reversal symmetric metal in region~$\gamma$. A common effect is an (approximately $\mathbf{k}$-independent) upward shift of the valence bands and a downward shift of the conduction bands, conventionally referred to as a scissor operator \cite{godby-GW-prb-1988,supp}. Meanwhile, the first conduction band is systematically flattened. In particular, for the Chern-insulating state, the band gap is reduced by about 65\% to 4.9\;meV, while the bandwidth decreases by about 25\%. This gap value is close to that inferred from a recent penetration-capacitance measurement in an hBN-aligned R5G moir\'e heterostructure ($\sim 3$\;meV) \cite{aronson-expR5GhBN-prx-2025}. More importantly, the quasiparticle weight of the low-energy bands for the Chern insulator state at $D=0.95$\;V/nm is around $0.9$, and throughout the entire phase diagram it never drops below $0.8$. This indicates that the R5G-hBN moir\'e system at integer fillings remains weakly correlated on top of the HF converged states.
    
    It is worth emphasizing that the dielectric constant $\epsilon_r$ is the only fitting parameter in our framework. Among $\epsilon_r=8-10$, the HF+RPA phase diagram obtained with $\epsilon_r=8$ shows the best overall agreement with the experimental phase diagram \cite{supp}, and $\epsilon_r=8$ is also consistent with recent experimentally measured value for graphene on hBN substrate \cite{lee-gra_QTM-arxiv-2025,yu-gra_capacitance-pnas-2013}. In addition, within the all-band HF approximations, we explored the possible emergence of an anomalous Hall crystal \cite{dong-R5GAHC-prl-2024,zhou-R5Gfci-prl-2024,dong-R5Gfci-prl-2024}. However, under experimentally relevant conditions of displacement field and dielectric constant, we find no $C=1$ gapped states at the HF level in the absence of moir\'e potential.
        
    \begin{figure}
        \centering
        \includegraphics[width=0.5\textwidth]{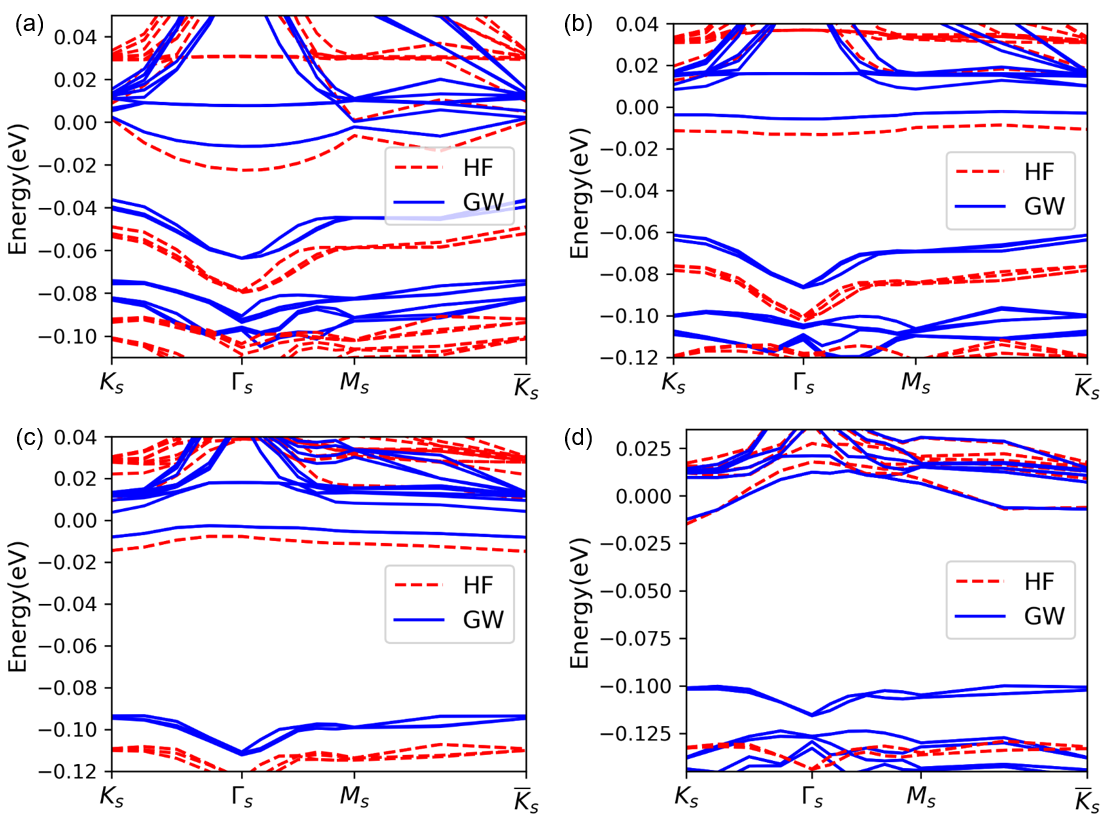}
        \caption{Comparison between $GW$ quasiparticle bands (blue solid lines) and HF bands (red dashed lines) using $\epsilon_r=8$ for the $\theta=0.77^{\circ}$ R5G-hBN heterostructure at $D=0.65$, 0.80, 0.95 and 1.00\;V/nm. These panels correspond, respectively, to (a) a time-reversal breaking metal and (b) a trivial insulator in region~$\alpha$, (c) a Chern insulator in region~$\beta$ and (d) a time-reversal symmetric metal in region~$\gamma$ \cite{data}. Fermi level is set to zero.}
        \label{fig:R5G_HF_GW_eps8}
    \end{figure}

    Next, we apply our techniques to MATBG described by a continuum model including both in-plane and out-of-plane lattice relaxation effects \cite{xie-arxiv-2025,supp}, which preserves $C_3$ symmetry and breaks particle-hole symmetry. With $\epsilon_r=8$, we find that, unlike most previous HF calculations \cite{kang-tbg-prl19,xie-tbg-prl20,zaletel-tbg-hf-prx20,jpliu-tbghf-prb21,cea-TBGHF-prb-2020}, the HF+$GW$+RPA ground state at $\nu=0$ is a nematic semimetal with two touching points near the moir\'e $\Gamma_s$ point, as seen from the HF and $GW$ band structures in Fig.~\ref{fig:TBG_nu0GS}(a). This state spontaneously breaks $C_3$ rotation symmetry \cite{shang-nematic-prr21,bultinck-tbg-strain-prl21} without the help of heterostrain, consistent with recent transport measurements \cite{cao-tbg-nematic-science21} and spectroscopic observations \cite{tbg-stm-andrei19,tbg-stm-caltech19,tbg-stm-pasupathy19,xiao-TBG_QTM-arxiv-2025}. The resulting $GW$ quasiparticle bands are mostly flat away from $\Gamma_s$, as observed in recent quantum twisting microscopy measurements \cite{xiao-TBG_QTM-arxiv-2025}. Compared to HF bands, the $GW$ correction reduces the flat-band width from 75\;meV to 47\;meV, in quantitative agreement with experiment ($\sim 50$\;meV) \cite{xiao-TBG_QTM-arxiv-2025}. The quasiparticle weight is around 0.9, indicating that the wavefunction of the ground state is close to a Slater determinant. The breaking of $C_3$ symmetry is clearly illustrated by the direct gap profile between the two lowest $GW$ bands shown in Fig.~\ref{fig:TBG_nu0GS}(c), with two minima approaching zero around $\Gamma_s$ point. By contrast, the K-IVC gapped state is found to be  metastable, as the nematic semimetal gains more correlation energy \cite{supp}.
    
    \begin{figure}
        \centering
        \includegraphics[width=0.48\textwidth]{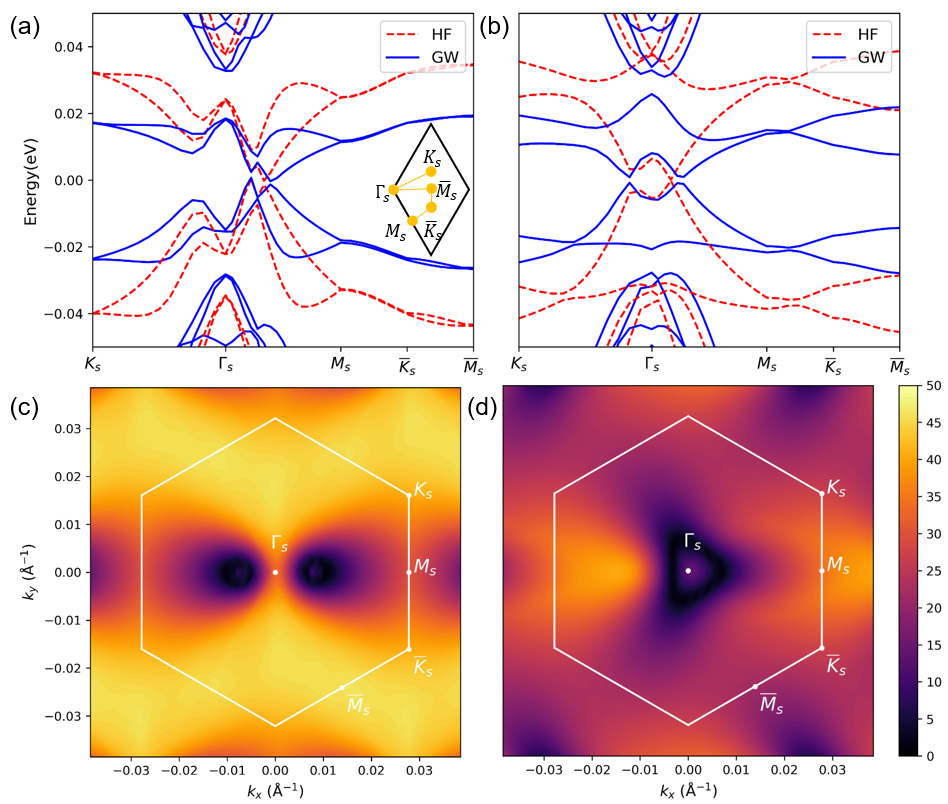}
        \caption{HF (red dashed) and $GW$ (blue solid) band structures for TBG of twist angle $\theta=1.08^\circ$ at $\nu=0$ for (a) nematic metal and (b) ``symmetric" metal, which share the same energy scale ($y$-axis). Bands plotted along a high-symmetry path, indicated in the inset of (a). (c,d) are the direct gaps (in meV) between the $GW$-corrected lowest conduction band and highest valence band for the nematic metal and symmetric metal, respectively.}
        \label{fig:TBG_nu0GS}
    \end{figure}

    Recently, several theoretical works tend to map MATBG onto Kondo-like problems using low-energy lattice models \cite{haule-mottTBG-arxiv-2019,carr-tbg_lattice-prr-2019}, topological heavy fermion model \cite{song-heavyfermion-prl22,shi-dai-heavy-prb22,Herzog-Arbeitman-iks_tbg-prb-2025}, and nonlocal moment description \cite{ledwith-nonlocalmoment_TBG-prx-2025,ledwith-tbg_diractrion-arxiv-2025},  with both analytical approaches \cite{hu-tbg_thf-arxiv-2025,ledwith-nonlocalmoment_TBG-prx-2025,ledwith-tbg_diractrion-arxiv-2025} and many-body numerical calculations \cite{haule-mottTBG-arxiv-2019,datta-DMFTtbg-natcomm-2023,rai-DMFTtbg-prx-2024,erez-berg-inprep1,erez-berg-inprep2}. These studies have proposed metallic ground states that seem like preserving rotational symmetry near the moir\'e $\Gamma_s$ at charge neutrality, dubbed ``symmetric metal", which competes with the K-IVC state in the absence of heterostrain, especially at finite temperatures. In our calculations, we also find a self-consistently converged ``symmetric metal" with a Mexican-hat-like dispersion near $\Gamma_s$, as shown in Fig.~\ref{fig:TBG_nu0GS}(b) by the HF and $GW$ band structures. In Fig.~\ref{fig:TBG_nu0GS}(d), we further show the direct gap profile extracted from the $GW$ bands of the ``symmetric metal" phase, which exhibits a ring-shaped minimum surrounding $\Gamma_s$ approximately preserving $C_3$. Away from $\Gamma_s$, however, $C_3$ symmetry is still broken. Our zero-temperature results place the competing states in ascending order of energy as follows: nematic semimetal, K-IVC state at 0.7\;meV per moir\'e unit-cell higher, and ``symmetric metal" at 14.4\;meV per moir\'e unit-cell higher. Results at other fillings \cite{supp} also show excellent agreement with experiment \cite{xiao-TBG_QTM-arxiv-2025} in both bandwidth and band shape. A detailed investigation of the ground states away from $\nu=0$ is left for our future work.

    In conclusion, we have developed a general beyond-mean-field framework for moir\'e materials that combines all-band HF approach with $GW$ quasiparticle corrections and RPA correlation energies, enabling a unified treatment of spontaneously symmetry-broken ground states in moir\'e systems. Applied to hBN-aligned R5G and MATBG, this approach substantially improves the agreement with experiments over band-projected HF approach, capturing the crucial role of dynamical screening and correlations in shaping phase competitions, quasiparticle dispersions, and excitation gaps. Our results highlight the importance of dynamical fluctuation effects in selecting the true ground state: in hBN-aligned R5G they produce a phase diagram in quantitative agreement with experiment, while in MATBG at charge neutrality they stabilize a nematic metal over the competing K-IVC state. More broadly, the many-body perturbative framework introduced here provides a systematic and quantitatively predictive route to interaction-driven phenomena in generic moir\'e systems.


\widetext
\clearpage

\begin{center}
\textbf{\large Supplemental Materials for ``General Many-Body Perturbation Framework for moir\'e Systems''} \\
\vspace{0.5cm}
Xin Lu, Yuanfan Yang, Zhongqing Guo, and Jianpeng Liu
\end{center}

\maketitle

\setcounter{equation}{0}
\setcounter{figure}{0}
\setcounter{table}{0}
\setcounter{page}{1}
\setcounter{section}{0}
\makeatletter
\renewcommand{\theequation}{S\arabic{equation}}
\renewcommand{\thesection}{S\arabic{section}}
\renewcommand{\thefigure}{S\arabic{figure}}
\renewcommand{\figurename}{Supplementary Figure}
\renewcommand{\tablename}{Supplementary Table}

\def\bibsection{\section*{References}} 
\tableofcontents

\section{S1. Continuum model}
\label{sec:continuum_model}

\subsection{Twisted hBN-aligned rhombohedral pentalayer graphene}
In our theoretical study, we adopt the continuum model derived by Moon and Koshino \cite{moon-hbn-graphene-prb14} to twisted pentalayer graphene-hBN (R5G-hBN) moir\'e superlattice, as we have done in Ref.~\cite{guo-HFfci-prb-2024}, where more details on the modelling can be found in the main text and the associated supplemental materials. The continuum model for R5G-hBN heterostructure is written as
\begin{align}
    H^{0,\mu} = 
    \begin{pmatrix}
        h^{0,\mu}_{\rm{intra}} + V_{\rm{hBN}} & (h^{0,\mu}_{\rm{inter}})^\dagger & 0 & 0 & 0 \\
        h^{0,\mu}_{\rm{inter}} & h^{0,\mu}_{\rm{intra}} & (h^{0,\mu}_{\rm{inter}})^\dagger & 0 & 0 \\
        0 & h^{0,\mu}_{\rm{inter}} & h^{0,\mu}_{\rm{intra}} & (h^{0,\mu}_{\rm{inter}})^\dagger & 0 \\
        0 & 0 & h^{0,\mu}_{\rm{inter}} & h^{0,\mu}_{\rm{intra}} & (h^{0,\mu}_{\rm{inter}})^\dagger \\
        0 & 0 & 0 & h^{0,\mu}_{\rm{inter}} & h^{0,\mu}_{\rm{intra}}
    \end{pmatrix}
\label{eq:Htot_all}
\end{align}
where $\mu=\pm 1$ is the valley index respectively for $\mathbf{K}_\mu$ ($\mathbf{K}\equiv \mathbf{K}_{+}$ and $\mathbf{K}' \equiv \mathbf{K}_{-}$). The intra- and inter-layer blocks are 
\begin{align}
    h^{0,\mu}_{\rm{intra}} &= - \hbar v_F^0 \k \cdot \bm{\sigma}_\mu \\
    h^{0,\mu}_{\rm{inter}} &=
    \begin{pmatrix}
       \hbar v_\perp (\mu k_x + i k_y) & t_\perp \\
        \hbar v_\perp (\mu k_x - i k_y) & \hbar v_\perp (\mu k_x + i k_y)
    \end{pmatrix}
\end{align}
where $\bm{\sigma}_\mu = (\mu \sigma_x, \sigma_y)$ are the Pauli matrices, representing sublattice $A/B$, and the value of the parameters are $\hbar v_F^0 = 5.253\,\eVA$, $\hbar v_\perp=0.335\,\eVA$ and $t_\perp = 0.34\,\eV$. In our study, we define the stacking geometry of the R5G-hBN system by starting from a non-rotated arrangement, where a $B/A$ site of graphene and a boron/nitrogen site of hBN share the same in-plane position, so that the in-plane $A$-$B$ bonds are parallel to each other. The effective moir\'e superlattice potential $V_{\rm{hBN}}$ acting directly on Layer 1 of pentalayer graphene, namely
\begin{align}
    V_{\rm{hBN}} = V^{\rm{eff}}(\br) + M^{\rm{eff}}(\br) \sigma_z + e v_F \mathbf{A}^{\rm{eff}}(\br) \cdot \bm{\sigma}_\mu.
\end{align}
where we classify different terms in the effective potential by their sublattice structure. Simple algebra calculations give
\begin{subequations}
    \begin{align}
        V^{\rm{eff}}(\br) &= V_0 - V_1 \sum_{j=1}^{3} \cos \alpha_j(\br) \\
        M^{\rm{eff}}(\br) &= \sqrt{3} V_1 \sum_{j=1}^{3} \sin \alpha_j(\br) \\
        e v_F \mathbf{A}^{\rm{eff}}(\br) &= 2 \mu V_1 \sum_{j=1}^{3} \begin{pmatrix}
            \cos [2\pi (j+1)/3] \\
            \sin [2\pi (j+1)/3]
        \end{pmatrix}\cos \alpha_j(\br) \\
        \alpha_j (\br) &=  \G_j \cdot \br + \psi + \frac{2 \pi}{3} \quad \rm{with} \quad \G_3 = -\G_1 - \G_2
    \end{align}
\end{subequations} 
where $V_0=0.0289\,\eV$, $V_1 = 0.0210\,\eV$ and $\psi = -0.29$\,rad. The moir\'e reciprocal vectors $\G_{1,2}$ form angle $120^\circ$ between them.

In multilayer graphene, an externally applied out-of-plane electric field is significantly screened due to the redistribution of electrons within different layers. This screening process is treated by solving the classical Poisson equation in electrostatics, while the charge density is calculated quantum mechanically using the continuum model. This is equivalent to making Hartree approximation to $e$-$e$ interactions assuming homogeneous in-plane charge density within each layer.

\subsection{Relaxed twisted bilayer graphene}
\begin{figure}
    \centering
    \includegraphics[width=0.8\textwidth]{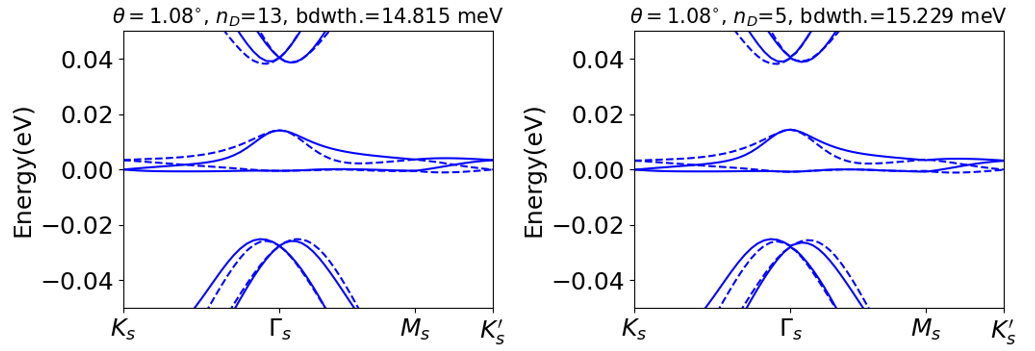}
    \caption{Non-interacting band structures for TBG at $\theta=1.08^{\circ}$ for two plane-wave cutoffs $n_D=13$ (left) and $n_D=5$ (right). The bandwidth of the flat bands are also given on the panels.}
    \label{fig:TBGnoninter}
\end{figure}

Based on Bistritzer-Macdonald continuum model for twisted bilayer graphene (TBG), we incorporate additionally lattice relaxation in our modelling, as we have derived in our recent work \cite{xie-arxiv-2025}. The relaxed lattice structure breaks particle-hole symmetry but preserves $C_{3z}$ rotation symmetry. The non-interacting band structures for twist angle $\theta=1.08^{\circ}$ are shown in Fig.~\ref{fig:TBGnoninter}, where we use two distinct plane-wave cutoffs $n_D$. The value of $n_D$ means we include $n_D^2$ plane-wave components, centered by the first Brillouin zone, in the continuum model. The low-energy bands are already converged within 0.5\;meV for $n_D=5$.  

\section{S2. All-band Hartree-Fock approximations}
In the all-band HF calculations, we consider only the dominant intravalley Coulomb interactions
\begin{equation}
    \begin{split}
&\hat{V}=\frac{1}{2N_s}\sum_{\substack{\alpha\alpha ', l l' \\\mu\mu ',\sigma\sigma '}}\sum_{\substack{\btk \btk ' \btq \\ \G \G' \Q}} V_{l l'}(\btq+\Q) \hat{c}^{\dagger}_{\sigma \mu l \alpha, \G+\Q}(\btk+\btq)  \\
&\times \hat{c}^{\dagger}_{\sigma' \mu ' l' \alpha ', \G'-\Q}(\btk ' - \btq) \hat{c}_{\sigma ' \mu ' l' \alpha ',\G'}(\btk ')\hat{c}_{\sigma \mu l \alpha,\G}(\btk)\;,
    \end{split}
\label{eq:V-intra}
\end{equation}
where $N_s$ is the number of moir\'e unit-cell and the annihilation (creation) operator $\hat{c}^{(\dagger)}_{\sigma \mu l \alpha,\G}(\btk)$ is associated with a plane wave component carry $\btk$ in the moir\'e Brillouin zone, indexed by moir\'e reciprocal vector $\G$, for electron with spin $\sigma$ belonging to valley $\mu$ at sublattice $\alpha$ of layer $l$. To model effectively the long-wavelength screening effects to the $e$-$e$ Coulomb interactions, we use a Coulomb interaction with Thomas-Fermi type of screening, whose Fourier transform is expressed as 
\begin{equation}
   V_{ll} (\mathbf{q})=\frac{e^2}{2 \Omega_0 \epsilon_r \epsilon_0 \sqrt{q^2+\kappa^2}}
\label{eq:V_thomasfermi}
\end{equation}
where $\Omega _0 = \sqrt{3}L_s^2 /2 $ is the area of the triangular moir\'e superlattice's primitive cell with moir\'e lattice constant $L_s$, $\epsilon_0$ the vacuum permittivity, $\epsilon_r$ the static homogeneous dielectric constant. We use the screening length $\kappa^{-1}=400$\;\AA. For the Coulomb interactions between electrons from different layers, we use
\begin{equation}
   V_{ll'}(\mathbf{q})=\frac{e^2}{2 \Omega_0 \epsilon_r \epsilon_0 q} e^{-q|l-l'|d_0}
   \label{eq:V_interlayer}
\end{equation}
with $l \neq l'$ and $d_0=3.35$\;\AA, the average distance between two adjacent layers. The divergence at $q=0$ should not be a concern, as it is physically regularized by the compensation from the positive charge background. This allows us to exclude the point $q=0$ from the calculations. A detailed formalism about the Hartree-Fock factorization and how to perform the subsequent self-consistent calculations can be found in our recent study \cite{guo-AHC-arxiv-2025} and its associated supplemental materials. Our results show that using different screening forms for the Coulomb potential, such as the double-gate form, does not affect the phase diagram.

In this study, we only consider layer-dependent Coulomb interactions for the calculations of R5G-hBN, but neglect such layer dependent screening for TBG. For both R5G-hBN and TBG, we use $n_D=5$. The $\bk$-mesh is $12 \times 12$ for R5G-hBN and $18 \times 18$ (at charge neutral $24 \times 24$) for TBG. We have verified numerically that the $\mathbf{k}$ point sampling and the plane-wave cutoff used in the all-band HF and $GW$ calculations are sufficient to ensure convergence. For magic-angle TBG, we have further checked that the results are robust against different choices of plane-wave-cutoff geometry, including the lozenge cutoff (given by $n_D$) and the $\mathbf{G}$-shell circular cutoff. For the lozenge cutoff, the results obtained with $n_D=5$ and 7, corresponding to 25 and 49 plane-wave components, respectively, are quantitatively consistent. For the circular cutoff, the results obtained with 3, 4, 5 $\mathbf{G}$ shells, corresponding to 19, 31, and 37 plane-wave components, respectively, are likewise quantitatively consistent.

It is worth noting that band-projected HF calculations cannot serve as a reliable starting point for many-body perturbation theory. Both the RPA correlation energy and the $GW$ self-energy are difficult to converge when the band space is truncated, which makes the all-band HF framework essential for the present analysis.

\subsection{Choice of initial conditions}
In our approach, the initial HF potentials are constructed systematically by enumerating all order parameters associated with relevant physical degrees of freedom, such as spin, valley, sublattice, and layer. This procedure allows us to access all HF-converged solutions, both ground states and metastable states, that can be characterized in terms of these order parameters. In this sense, when the characterization of HF states is restricted to order parameters defined by physical degrees of freedom, the procedure can be regarded as an unconstrained HF calculation. In the HF+RPA framework, we then compute the RPA correlation energy for all HF-converged states and determine the ground state by combining it with the HF total energy.

In the present all-band HF calculations, it is convenient to specify the initial conditions in the original basis, where all physical degrees of freedom are explicit. Since our focus is on states characterized by spin, valley, and sublattice order, the initial HF potential is taken to be proportional to the tensor product of Pauli matrices $s_\alpha \tau_\beta \sigma_\gamma$ where $s , \tau , \sigma$ represents spin, valley and sublattice degrees of freedom, respectively, and $\alpha,\beta,\gamma=0,x,y,z$. The amplitude of the initial potential is chosen to be 0.1 eV, and it is taken to be momentum independent across the moir\'e Brillouin zone. The latter is not restrictive, as the momentum dependence of the HF potential is generated self-consistently during the HF iterations.

\subsection{Bare Hartree-Fock Phase diagrams for R5G-hBN}
\begin{figure}
    \centering
    \includegraphics[width=0.8\textwidth]{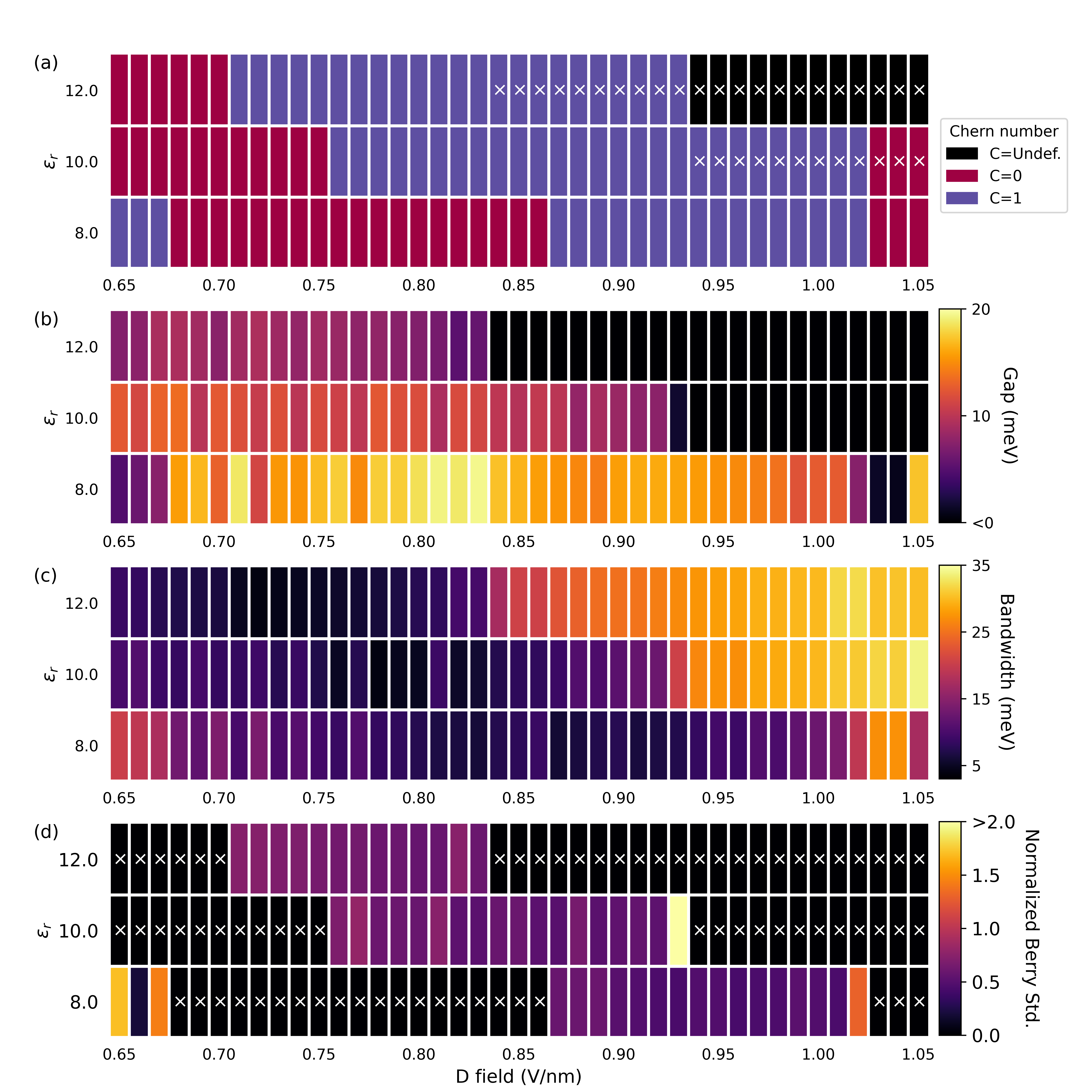}
    \caption{Phase diagram of bare all-band HF results for R5G-hBN including (a) Chern number, (b) indirect gap, (c) bandwidth and (d) normalized Berry curvature of the first conduction band. Chern number is undefined if direct gap is vanishing. The white cross indicates the metallic state in (a) while it rules out the state with topologically trivial state or metallic state in (d).}
    \label{fig:PhD_HFonly_all_eps8-12}
\end{figure}

\begin{figure}
    \centering
    \includegraphics[width=0.5\textwidth]{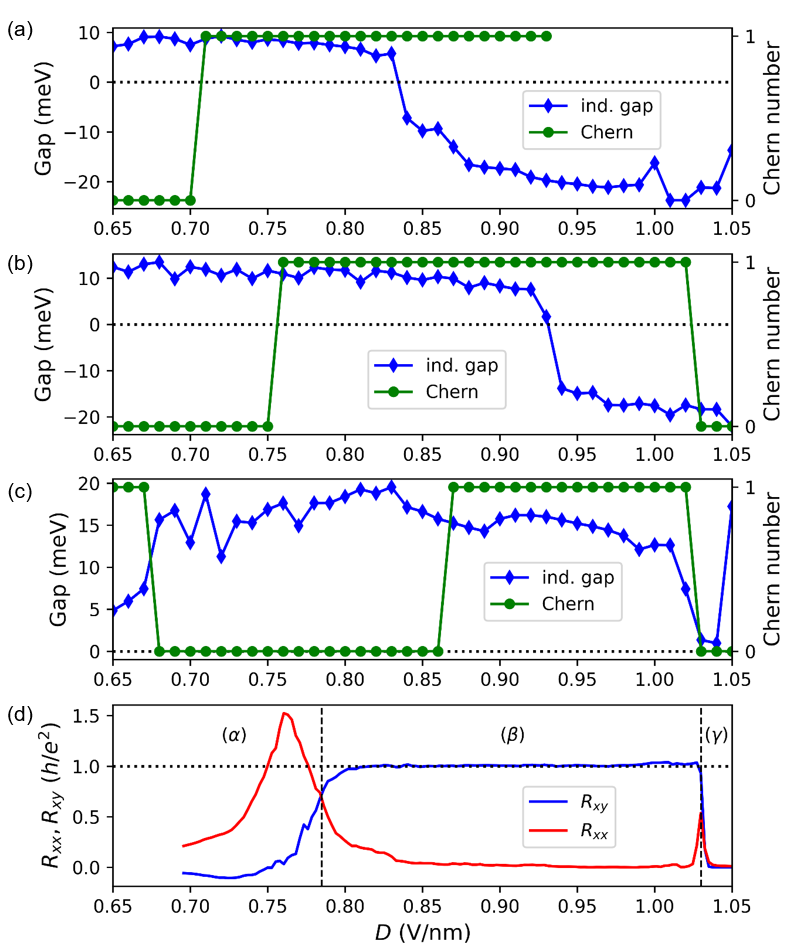}
    \caption{Variation of the Chern number of the first conduction band, that remains separated from the second conduction band by a finite direct gap, together with the indirect gap, as a function of the displacement field $D$ for $\epsilon_r=8,10,12$ in (a,b,c), respectively, for $\theta=0.77^\circ$ R5G-hBN using bare all-band HF. The experimental results are shown in (d) for comparison.}
    \label{fig:PhD_HFonly_Chern_indGap_eps8-12}
\end{figure}

We present the detailed bare HF phase diagrams for $\theta=0.77^\circ$ R5G-hBN at filling $\nu=1$ under displacement fields $D$ ranging from 0.65 to 1.05\;V/nm in Fig.~\ref{fig:PhD_HFonly_all_eps8-12} and Fig.~\ref{fig:PhD_HFonly_Chern_indGap_eps8-12}. In Fig.~\ref{fig:PhD_HFonly_all_eps8-12}, the Chern number, indirect gap, bandwidth, and normalized Berry curvature of the first conduction band are shown in color scale. The corresponding numerical values of the Chern number and indirect gap are presented in Fig.~\ref{fig:PhD_HFonly_Chern_indGap_eps8-12}.

Among the three dielectric constants considered, $\epsilon_r=8,10,$ and $12$, the bare HF phase diagram with $\epsilon_r=10$ provides the best qualitative agreement with experiment, as it reproduces the observed evolution of the ground state with increasing displacement field: from metal to trivial insulator, then to Chern insulator, and finally back to a metallic phase. Although the phase diagram for $\epsilon_r=12$ shows a similar sequence of phases, it underestimates the extent of the Chern insulating region compared with the $\epsilon_r=10$ case. Nevertheless, all bare HF phase diagrams tend to overestimate the stability of the trivial insulating state.

\section{S3. Random phase approximation for correlation energy}

The total energy of the system within RPA framework is given by:
\begin{equation}
	E_{\text{tot.}}=E_{\text{kin.}}+E_{\text{HF}}+E_{c}^{\text{RPA}},
\end{equation}
where $E_{\text{kin.}}$ is the kinetic energy, $E_{\text{HF}}$ is the Hartree-Fock energy, and $E_{c}^{\text{RPA}}$ represents the RPA correlation energy, which is given by \cite{fetter-book,bohm-rpa-pr53,gellmann-rpa-pr57,ren-rpa-2012}:

\begin{equation}
	\begin{aligned}
		E_c^{\text{RPA}}&=\int_{-\infty}^{\infty} \frac{\mathrm{d} \omega}{4 \pi}  \sum_{\btq,\Q,\Q'} \left\{ \ln \left[\delta_{\Q\Q'} - V_{\Q } (\btq)\delta_{\Q\Q'}\chi^0_{\Q'\Q }(\btq,i\omega)\right] \right. \\
        &\hspace{3cm} \left. + V_{\Q }(\btq)\delta_{\Q\Q'}\chi^0_{\Q'\Q }(\btq,i\omega) \right\},
	\end{aligned}
    \label{eq:EcRPA}
\end{equation}
where $V_{\Q } (\btq)=V(\Q+\btq)$ is the bare Coulomb interaction, and $\chi^0(i\omega)$ is the bare charge polarizability in the imaginary frequency domain calculated from HF wavefunctions and single-particle spectra. $\btq$ and $\Q$ denote wave vectors, and $\chi^0_{\Q'\Q}(\btq,i\omega)$ are the matrix elements of the bare charge susceptibility in the original plane-wave basis. The matrix elements in reciprocal space are
\begin{equation}
	\begin{aligned}
		\chi^{0}_{\bQ\bQ'}(\btq,\nu)=&\frac{1}{N_s\Omega_0}\sum_{n',n,\btk}
		\left[\sum_{\lambda,\bG}
		C^*_{\lambda\bG+\bQ, n'\btk+\btq}
		C_{\lambda\bG, n\btk}\right]^*
		\left[\sum_{\lambda',\bG'}
		C^*_{\lambda'\bG'+\bQ', n'\btk+\btq}
		C_{\lambda'\bG', n\btk}\right] \\
		&\times\left[\frac{\theta(\varepsilon_{n'\btk+\btq}-\varepsilon_F)\theta(\varepsilon_F-\varepsilon_{n\btk})}{\nu-\varepsilon_{n'\btk+\btq}+\varepsilon_{n\btk}+i\delta}
		-\frac{\theta(\varepsilon_F-\varepsilon_{n'\btk+\btq})\theta(\varepsilon_{n\btk}-\varepsilon_F)}{\nu-\varepsilon_{n'\btk+\btq}+\varepsilon_{n\btk}-i\delta}\right],
        \label{eq:chi0}
	\end{aligned}
\end{equation}
where \( C_{\lambda\bG, n\btk} \) are the expansion coefficients of the single-particle states in the plane-wave basis, $\varepsilon_{n\btk}$ are the HF eigenvalues, and $\varepsilon_F$ is the Fermi energy, and $\bG$, $\bG'$, $\bQ$, and $\bQ'$ denote moir\'e reciprocal vectors.

\section{S4. $GW$ approximations}

The key equations in the $GW$ formalism are known as Hedin's equations, which describe a set of self-consistent equations of self-energy $\Sigma$, Green's function $G$, vertex function $\Gamma$, polarization propagator $P$, and screened Coulomb interaction $W$. For simplicity, here we use Arabic number as a short-hand notation for spatial ($\br$) and temporal ($t$) coordinate, e.g., $1\equiv(\br_1, t_1)$. Then, with such short-hand notations, Hedin's equations are given by:
\begin{equation}
	\Sigma(12) = i \int \text{d}3 \, G(13) W(14) \Gamma(342),
\end{equation}
\begin{equation}
	G(12) = G_0(12) + \int \text{d}3 \, G_0(13) \Sigma(34) G(42),
\end{equation}
\begin{equation}
	\Gamma(123) = \delta(12) \delta(13) + \int \text{d}4 \, \text{d}5 \, \frac{\delta \Sigma(12)}{\delta G(45)} G(46) G(75) \Gamma(673),
\end{equation}
\begin{equation}
	P(12) = -i \int \text{d}3 \, \text{d}4 \, G(13) G(42) \Gamma(342),
\end{equation}
\begin{equation}
	W(12) = V(12) + \int \text{d}3 \, V(13) P(34) W(42).
\end{equation}

These equations form the basis of many-body perturbation theory, linking the Green's function, self-energy, and screened interaction in a self-consistent framework. To apply the $GW$ approximation, we assume a simple form of vertex function, $\Gamma(123) = \delta(12) \delta(13)$, which leads to the ``bare vertex'' approximation:
\begin{equation}
	\Gamma(123) = \delta(12) \delta(13).
\end{equation}
Under this approximation, the self-energy simplifies to:
\begin{equation}
	\Sigma(12) = i G(12) W(12).
\end{equation}
Similarly, the polarization reduces to:
\begin{equation}
	P(12) = -i G(12) G(21).
\end{equation}
To describe the screened Coulomb interaction $W$ in terms of the dielectric function $\epsilon$, we express $V$ as:
\begin{equation}
	V(12)=\int\text{d}{3}\epsilon(13)W(32).
\end{equation}
The dielectric function $\epsilon$ can be formulated as:
\begin{equation}
	\epsilon(12)=\delta(12)-\int\text{d}{3}V(13)P(32).
\end{equation}

We continue to perform a Fourier transform from time domain to frequency domain to handle the time-dependent components. Now we go back to the usual notations where $\br$ ($\br'$) denote real-space coordinate, and $\omega$ denote frequency. The bare Green's function $G_0$ is given by:
\begin{align}
	G_0(\br,\br',\omega)=\sum_{n\btk}\frac{\psi_{n\btk}(\br)\psi^*_{n\btk}(\br')}{\omega-\varepsilon_{n\btk}+i\delta\mathrm{sgn}(\varepsilon_{n\btk}-\varepsilon_F)},
\end{align}
where $\psi_{n\btk}$ are the HF single-particle wave functions, $\varepsilon_{n\btk}$ are the HF eigenvalues, and $\varepsilon_F$ is the Fermi energy. The bare charge polarizability $\chi^0$, which characterizes the linear response of the system to external perturbative potentials, is given by:
\begin{equation}
	\chi^{0}(\br,\br',\nu)=-i\int\frac{\text{d}{\omega}}{2\pi}e^{i\omega\delta^+}G_0(\br,\br',\omega+\nu)G_0(\br',\br,\omega).
\end{equation}
Using $\chi^0$, the dielectric function within the Random Phase Approximation (RPA) is expressed as:
\begin{equation}
	\epsilon_{\text{RPA}}(\br, \br', \omega)=\delta(\br, \br')-\int\text{d}\br''V(\br, \br'')\chi^{0}(\br'', \br',\omega).
\end{equation}
The screened Coulomb interaction within RPA is:
\begin{equation}
	W_{\text{RPA}}(\br, \br', \omega) = \int \text{d}\br'' \left[ \epsilon^{-1}_{\text{RPA}}(\br, \br'', \omega) - \delta(\br, \br'') \right] V(\br'', \br'),
\end{equation}
where the static part has been subtracted as it is already taken into account in the HF calculations.
The correlation part of the self-energy, $\Sigma_c$, which accounts for electron correlation effects beyond HF, is then computed using:
\begin{equation}
	\Sigma_c(\br, \br', \omega) = \frac{i}{2\pi} \int d\nu \, e^{i\nu\delta^+} G_0(\br, \br', \omega + \nu) W_{\text{RPA}}(\br', \br, \nu),
\end{equation}
where $G_0$ is the HF Green's function and $W_{\text{RPA}}$ is the dynamically screened Coulomb interaction.

Then, we perform a Fourier transform from real space to reciprocal space. The matrix elements of the bare charge polarizability in reciprocal space are given by Eq.~\eqref{eq:chi0}. The matrix form of the RPA dielectric function in reciprocal space is expressed as:
\begin{equation}
	\epsilon^{\text{RPA}}_{\bQ\bQ'}(\btq,\omega)	=\delta_{\bQ\bQ'}-V(\btq+\bQ)\chi^{0}_{\bQ\bQ'}(\btq,\omega),
\end{equation}
and the screened Coulomb interaction in reciprocal space is given by:
\begin{equation}
	W^{\text{RPA}}_{\bQ\bQ'}(\btq,\omega)=\left[ \epsilon^{-1,\text{RPA}}_{\bQ\bQ'} (\btq,\omega)-\delta_{\bQ\bQ'}\right]V(\btq+\bQ').
\end{equation}

The correlation self-energy in reciprocal space can be expressed as:
\begin{equation}
	\begin{aligned}
		\Sigma_{c}(\btk,\omega)_{n' n}
		=&\frac{i}{N_s\Omega_0}\sum_{m,\btq}\sum_{\bG,\bG'}
		\left[ \sum_{\lambda,\bQ}C^*_{\lambda\bG+\bQ,m\btk+\btq}C_{\lambda\bQ,n'\btk}\right] ^*
		\left[ \sum_{\lambda',\bQ'}C^*_{\lambda'\bG'+\bQ',m\btk+\btq}C_{\lambda'\bQ',n\btk}\right]  \nonumber\\
		&\times V(\btq+\bG)
		\int\frac{\text{d}{\nu}}{2\pi}e^{i\nu\eta}
		\frac{\left[ \epsilon^{-1,\text{RPA}}_{\bG'\bG}(\btq,\nu)-\delta_{\bG\bG'}\right] }{\omega+\nu-\varepsilon_{m\btk+\btq}+i\delta\mathrm{sgn}(\varepsilon_{m\btk+\btq}-\varepsilon_F)} ,
	\end{aligned}
\end{equation}
where the notations follow those in Eq.~\eqref{eq:chi0}.

\subsection{Compare different $GW$ schemes}
Our implementation of $GW$ approximations consider only the diagonal part of self-energy. The quasiparticle energies can be then obtained using linear expansion in self-energy
\begin{equation}
	\varepsilon_{n\btk}^{\text{QP}} = \varepsilon_{n\btk}^{\text{HF}} + Z_{n\btk} \, \text{Re} \, \Sigma_c(\btk, \varepsilon_{n\btk}^{\text{HF}})_{nn},
    \label{eq:qp_diagonal}
\end{equation}
where $Z_{n\btk}$ is the quasiparticle weight, accounting for interaction renormalization effects of quasiparticles
\begin{equation}
	Z_{n\btk} = \left[ 1 - \text{Re} \left( \frac{\partial \Sigma_c(\btk, \omega)_{nn}}{\partial \omega} \right)_{\omega = \varepsilon_{n\btk}^{\text{HF}}} \right]^{-1}.
\end{equation}
A one-shot $GW$ calculation, commonly referred to as the $G_0W_0$ scheme \cite{aryasetiawan-gw-rpp98,hybertsen-GW0-prb-1986}, often provides already significant corrections to quasiparticle energies. A relatively inexpensive improvement is to update the HF energies in Eq.~\eqref{eq:qp_diagonal} with the quasiparticle energies and iterate until convergence. This eigenvalue-only $GW$ scheme (EV-$GW$) \cite{surh-EVGW-prb-1991}, which is what we use to obtain the figures in the main text, systematically improves the accuracy of quasiparticle energies compared to $G_0W_0$, especially when the initial HF energies are far from the final quasiparticle energies. 

The off-diagonal part of the self-energy can be included by taking its Hermitian component, as done in the quasiparticle self-consistent $GW$ ($QSGW$) scheme \cite{schilfgaarde-QPGW-prl-2006,kotani-QPGW-prb-2007}. Previous studies indicate that gaps obtained by different $GW$ schemes differ by at most 10\% \cite{hybertsen-GW0-prb-1986,hybertsen-GW-prl-1985}, which corresponds to 1\;meV in our case. Our calculations given in Table~\ref{tab:GW} show consistent results across different $GW$ schemes without significant discrepancies \cite{rodl-vertex_GW-prb-2015,bruneval-compareGW-prb-2006}. Therefore, in our theory, we adopt the EV-$GW$ scheme, calculating $GW$ quasiparticle bands separately, as we only consider the diagonal part of the self-energy.

\begin{table}[ht]
    \centering
    \begin{tabular}{ccccc}
    \hline
    $GW$ scheme (meV) & HF & $G_0 W_0$ & EV-$GW$ & $QSGW$ \\
    \hline \hline
    Indirect gap     & 11.8      & 3.4         & 3.9          & 3.2        \\
    \hline
    Direct gap      & 16.6      & 6.2          & 6.7           & 5.6        \\
    \hline
    Bandwidth        & 4.8        & 3.6        & 3.6           & 3.1         \\
    \hline    
\end{tabular}
    \caption{Comparison between HF results and $GW$ results using different schemes for $\epsilon_r=10$ at $D=0.8$\;V/nm in R5G-hBN case.}
    \label{tab:GW}
\end{table}

\begin{figure}
    \centering
    \includegraphics[width=0.8\textwidth]{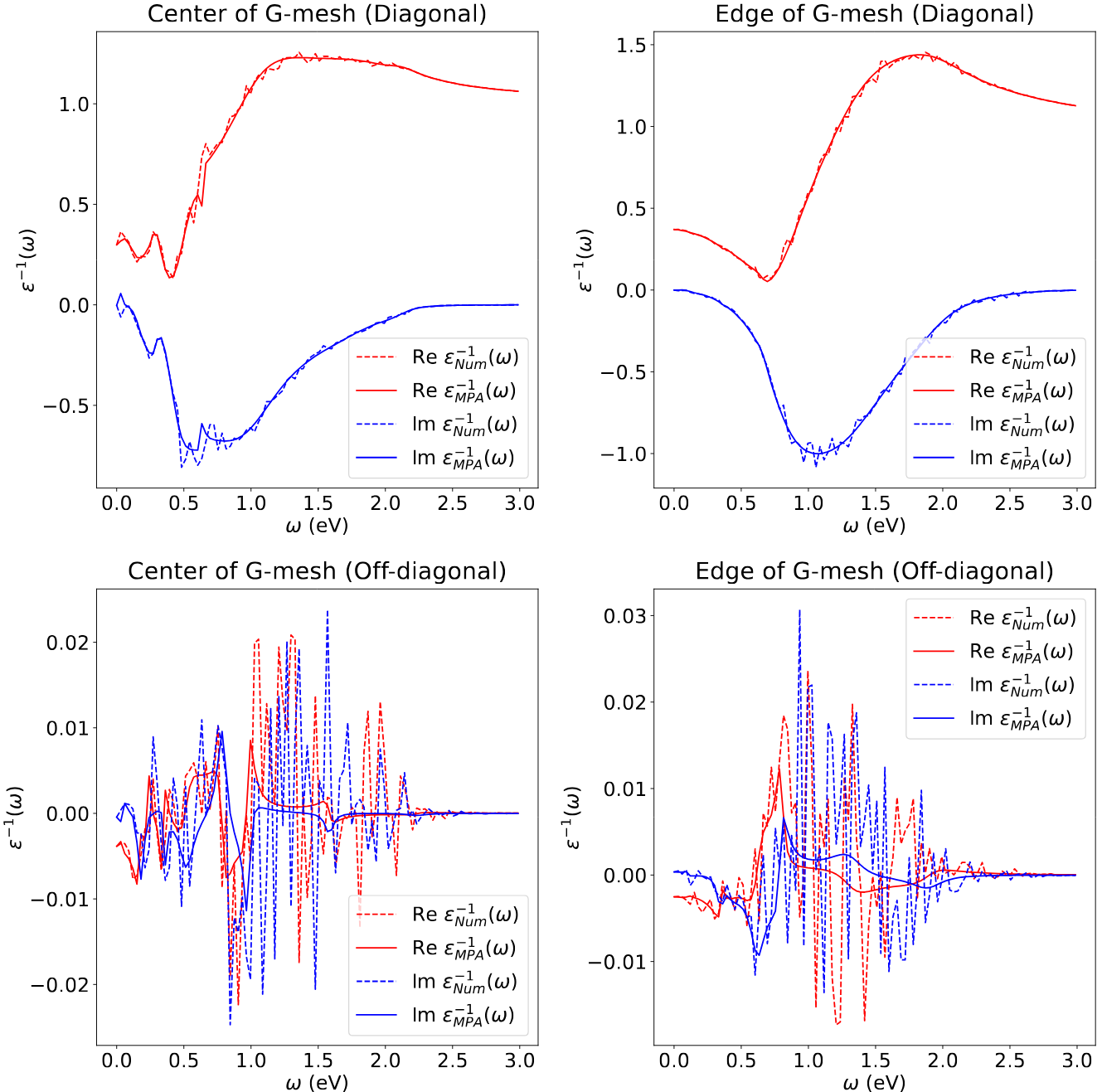}
    \caption{Comparison between the numerically calculated inverse dielectric function (dashed lines) and the MPA-fitted inverse dielectric function (solid lines). The real part is shown in red, and the imaginary part is shown in blue. Left upper panel: small $\q$ with $\G = 0$ near the $\Gamma$ point; right upper panel: non-zero $\G$ away from the $\Gamma$ point; two lower panels: off-diagonal elements near (left) and away from (right) the $\Gamma$ point.}
    \label{fig:MPAfit_smallq}
\end{figure}

\begin{figure}
    \centering
    \includegraphics[width=0.8\textwidth]{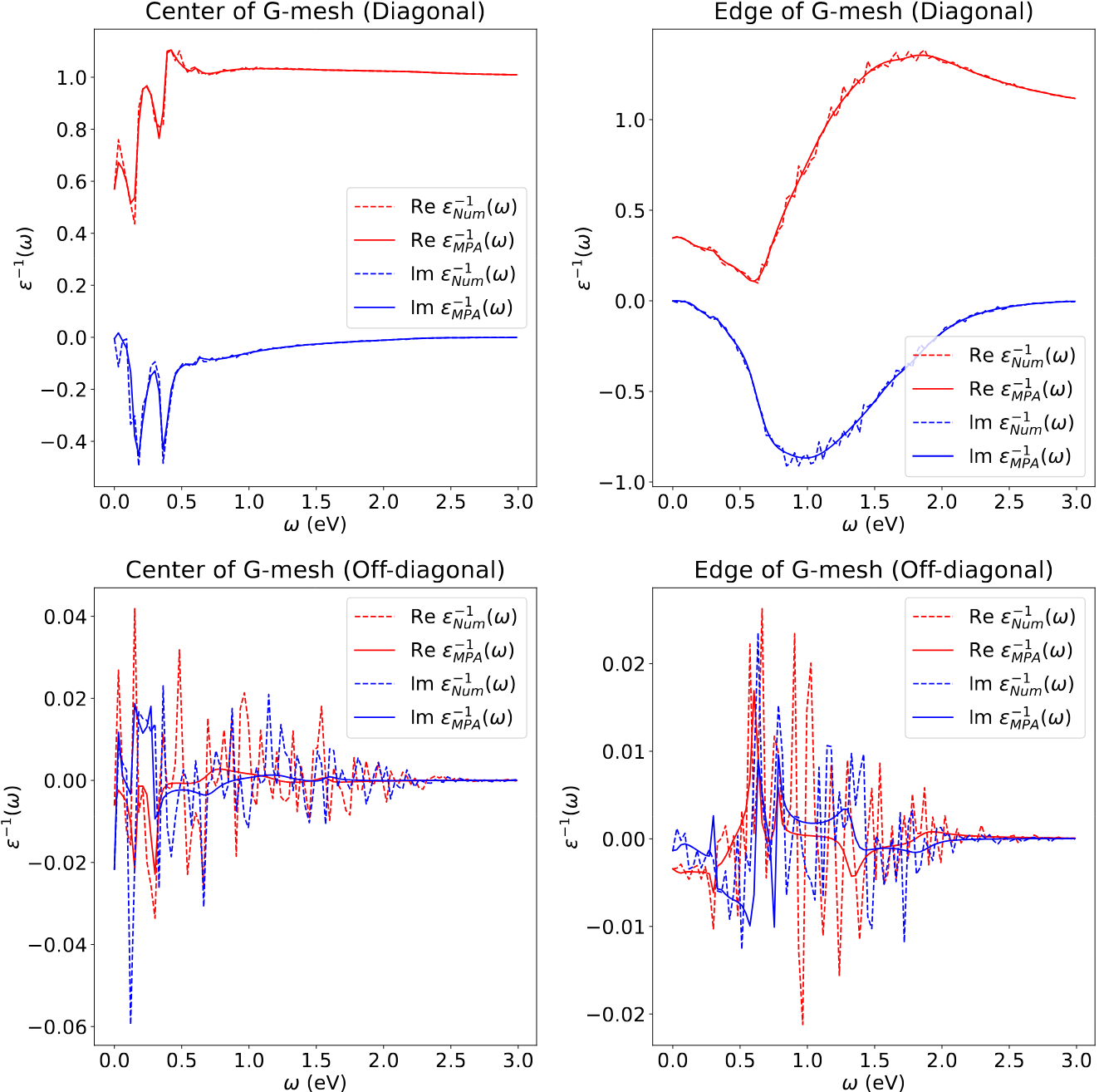}
    \caption{Same as Fig.~\ref{fig:MPAfit_smallq} except for large $\q$.}
    \label{fig:MPAfit_largeq}
\end{figure}

At this stage, advanced $GW$ schemes are not crucial for several reasons. First, more complex methods do not necessarily improve the results, as cancellation effects between various vertex corrections can occur \cite{kotani-QPGW-prb-2007, bobbert-vertex_GW-prb-1994, kutepov-vertex_GW-prb-2016}. Consequently, such methods often yield results that are comparable to those from $G_0 W_0$ \cite{delsole-GWGamma-prb-1994, mahan-GWGamma-prl-1989}. Additionally, the off-diagonal part of the self-energy is usually less important, as shown in Fig.~\ref{fig:MPAfit_smallq} and Fig.~\ref{fig:MPAfit_largeq}, since the converged wavefunctions typically resemble the HF wavefunctions. 

\subsection{Multiple plasmon pole approximation}
The most numerically demanding part of the EV-$GW$ method is the calculation of the polarizability, which involves a double summation over bands at multiple frequencies to compute the self-energy. A naive evaluation of the self-energy by computing $\chi^0$ at many frequencies to ensure convergence would be extremely computationally demanding, especially in self-consistent EV-$GW$ loops where quasiparticle energies must be updated both in $G$ and in $\chi$ (hidden in $W$). This motivates the use of an approximation that simplifies the numerical evaluation of the self-energy: the multiple plasmon-pole approximation (MPA) \cite{mpa-prb21,mpa-prb23}. 

The full derivation of the MPA for the continuum model is given in literature \cite{mpa-prb21,mpa-prb23}. Our theory adapts their formalism to moir\'e systems. The dielectric function is then fitted using these multiple plasma (collective-excitation) modes. The approximation is given by:
\begin{equation}
	\epsilon^{-1,MPA}_{\bQ\bQ'}(\btq,\nu)-\delta_{\bQ\bQ'}
	=\sum^{N_p}_{l}\frac{2R_{l,\bQ\bQ'}(\btq)\Omega_{l,\bQ\bQ'}(\btq)}{\nu^2-\Omega^2_{l,\bQ\bQ'}(\btq)},
    \label{eq:MPA}
\end{equation}
where $N_p$ represents the number of plasmon poles, while $R$ and $\Omega$ are parameters to be determined. Specifically, $R_{l,\bQ\bQ'}(\btq)$ denotes the residue of the $l$-th plasmon mode, and $\Omega_{l,\bQ\bQ'}(\btq)$ represents its frequency. To solve for these unknown parameters, we require the values of the dielectric function at $2N_p$ different frequencies. We set $N_p=10$ in this work. 

Using the MPA, we can compute the correlation self-energy more efficiently without losing accuracy. The correlation self-energy $\Sigma_c$ in the MPA is given by:
\begin{align}
		\Sigma_{c}(\btk,\omega)_{nn}=&\frac{1}{N_s\Omega_0}\sum_{m,\btq}\sum_{\bG,\bG'}
		\left[ \sum_{\lambda,\bQ}C^*_{\lambda\bG+\bQ,m\btk+\btq}C_{\lambda\bQ,n\btk}\right] ^*
		\left[ \sum_{\lambda',\bQ'}C^*_{\lambda'\bG'+\bQ',m\btk+\btq}C_{\lambda'\bQ',n\btk}\right]  \nonumber\\
		\times &\sum^{N_p}_{l}\frac{V(\btq+\bG)R_{\bG'\bG,l}(\btq)}{\omega-\varepsilon_{m\btk+\btq}+i\delta\mathrm{sgn}(\varepsilon_{m\btk+\btq}-\varepsilon_F)+\Omega_{\bG'\bG,l}(\btq)(2f_{m\btk+\btq}-1)},
	\label{eq:self-energy}
\end{align}
where $f_{m\btk+\btq}$ represents the Fermi-Dirac distribution function.
The derivative of the correlation self-energy with respect to $\omega$ is required to determine the QP weight $Z_{n\btk}$:
\begin{align}
		\frac{\partial \Sigma_{c}(\btk,\omega)_{nn}}{\partial \omega}=&\frac{-1}{N_s\Omega_0}\sum_{m,\btq}\sum_{\bG,\bG'}
		\left[ \sum_{\lambda,\bQ}C^*_{\lambda\bG+\bQ,m\btk+\btq}C_{\lambda\bQ,n\btk}\right] ^*
		\left[ \sum_{\lambda',\bQ'}C^*_{\lambda'\bG'+\bQ',m\btk+\btq}C_{\lambda'\bQ',n\btk}\right]  \nonumber\\
		\times &\sum^{N_p}_{l}\frac{V(\btq+\bG)R_{\bG'\bG,l}(\btq)}{\left[ \omega-\varepsilon_{m\btk+\btq}+i\delta\mathrm{sgn}(\varepsilon_{m\btk+\btq}-\varepsilon_F)+\Omega_{\bG'\bG,l}(\btq)(2f_{m\btk+\btq}-1)\right] ^2}.
	\label{eq:self-energy-diff}
\end{align}

Fig.~\ref{fig:MPAfit_smallq} (small $\q$ momentum transfer) and Fig.~\ref{fig:MPAfit_largeq} (large $q$ momentum transfer) illustrate a comparison between the numerically calculated inverse dielectric function (dashed lines) and the MPA-fitted inverse dielectric function (solid lines), based on the HF results for R5G-hBN using $D=0.8$\;V/nm and $\epsilon_r=10$. The red and blue lines correspond to the real and imaginary parts, respectively. The MPA successfully represents the continuous spectrum using multiple plasmons. Additionally, oscillations in the numerically computed inverse dielectric function, arising from $\bk$-mesh discretization or finite-size effects, are smoothed out by the MPA. These small discrepancies have negligible impact on the final self-energy integral. Although these elements are relatively small, the MPA still provides an accurate description.

\section{S5. More results for hBN-aligned R5G}

\subsection{Phase boundary for HF+RPA phase diagram}
For the boundary between region $\beta$ and $\gamma$, the calculations with $\epsilon_r=8$ yields 1.02\;V/nm, whereas those using $\epsilon_r=9$ and 10 give 0.89 and 0.94\;V/nm, respectively, both deviating more substantially from the experimental value of 1.03\;V/nm.

\begin{figure}
    \centering
    \includegraphics[width=0.8\textwidth]{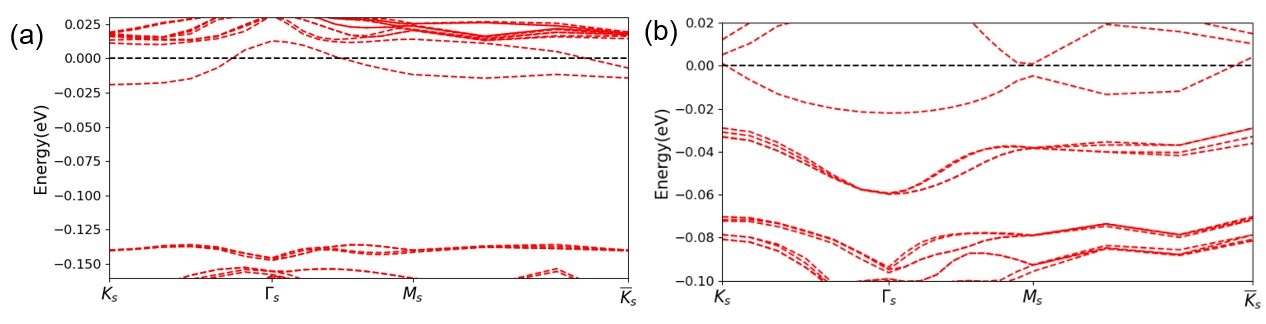}
    \caption{Typical HF single-particle spectra for (a) the state at high $D$ field ($D\sim 1.0$\;V/nm) and (b) that at low field $D$ ($D\sim0.6$\;V/nm) in the phase diagram of $\epsilon_r=10$. Both phases are metallic.}
    \label{fig:R5G_metallic}
\end{figure}

\subsection{Two metallic states}

In this section, we show the difference between two metallic phases having a $C=1$ first conduction band that forms a finite direct gap with the second conduction band in the phase diagram of $\epsilon_r=10$. The low-field metallic state is the same metallic state given in Fig.~3(a) in the main text for $\epsilon_r=8$. In Fig.~\ref{fig:R5G_metallic}, we show typical HF single-particle spectra for the state at high $D$ field ($D\sim 1.0$\;V/nm) and that at low field $D$ ($D\sim0.6$\;V/nm).

The Fermi surface of the high-field state consists of a small hole pocket around the moir\'e $\Gamma_s$ point from the first conduction band and a small electron pocket from the second conduction band around the moir\'e $K_s$ point. When we compute the total Berry curvature below the Fermi surface, we obtain finite values, which can give rise to an anomalous Hall resistance unlike a degenerate metal state.

In contrast, the Fermi surface of the low-field state contains hole pockets at the moir\'e $K_s$ points of the first conduction band and electron pockets at the moir\'e $M_s$ point of the second conduction band. All of these regions carry substantial Berry curvature. As a result, the balance between the Berry-curvature contributions from the hole and electron pockets strongly affects the total Berry curvature below the Fermi level and therefore the Hall resistance, which can deviate significantly from the resistance quantum, consistent with experimental observations.

\subsection{Compare HF and $GW$ bands}
A more detailed comparison between the HF band structure and the $GW$ quasiparticle band structure for the ground states of the $\theta=0.77^\circ$ R5G-hBN heterostructure is given in Table~\ref{Tab:compareHFGW}. We highlight the difference on the band gaps and the bandwidth of the first conduction band for $D=0.65$, 0.80, 0.95 and 1.00\;V/nm with $\epsilon_r=8$.
\begin{table}[!htbp]
    \caption{Comparison between the HF band structure and the $GW$ quasiparticle band structure for the ground states of the $\theta=0.77^\circ$ R5G-hBN heterostructure at $D=0.65$, 0.80, 0.95 and 1.00\;V/nm with $\epsilon_r=8$, focusing on the band gaps and the bandwidth of the first conduction band. Energies are in meV.}
    \label{Tab:compareHFGW}
    \begin{ruledtabular}   
    \begin{tabular}{c| c c c c}
    $D$ (V/nm) & HF/GW & direct gap & indirect gap & bdwth. \\
    \hline
    0.65& HF & 1.6 & -2.7 & 23.9\\ 
    0.65& GW & 0.4 & -3.0 & 13.8\\ 
    0.80& HF & 20.6 & 18.4 & 7.5\\ 
    0.80& GW & 10.1 & 7.8 & 5.4\\ 
    0.95& HF & 21.5 & 15.2 & 9.5\\ 
    0.95& GW & 11.3 & 4.9 & 7.1\\ 
    1.00& HF & 0.0 & -36.4 & 36.4\\ 
    1.00& GW & 0.0 & -28.3 & 28.3 
    \end{tabular}
    \end{ruledtabular}
\end{table}

\section{S6. More results for magic-angle TBG}
In this section, we show more results for TBG, in particular: 
\begin{enumerate}
    \item HF and $GW$ band structures of the ground state for TBG of $\theta=1.08^\circ$ at $\nu=\pm 2$, determined by HF+$GW$+RPA technique in which 24 $GW$ bands are included.  
    \item HF and $GW$ band structures of the metastable state gapped at CNP for TBG of $\theta=1.08^\circ$ at $\nu=0$  
    \item HF and $GW$ band structures of the ground state and two metastable states (determined by HF+$GW$+RPA technique in which 24 $GW$ bands are included) for TBG of $\theta=1.08^\circ$ at $\nu=-0.2$  
\end{enumerate}

\begin{figure}
    \centering
    \includegraphics[width=0.5\textwidth]{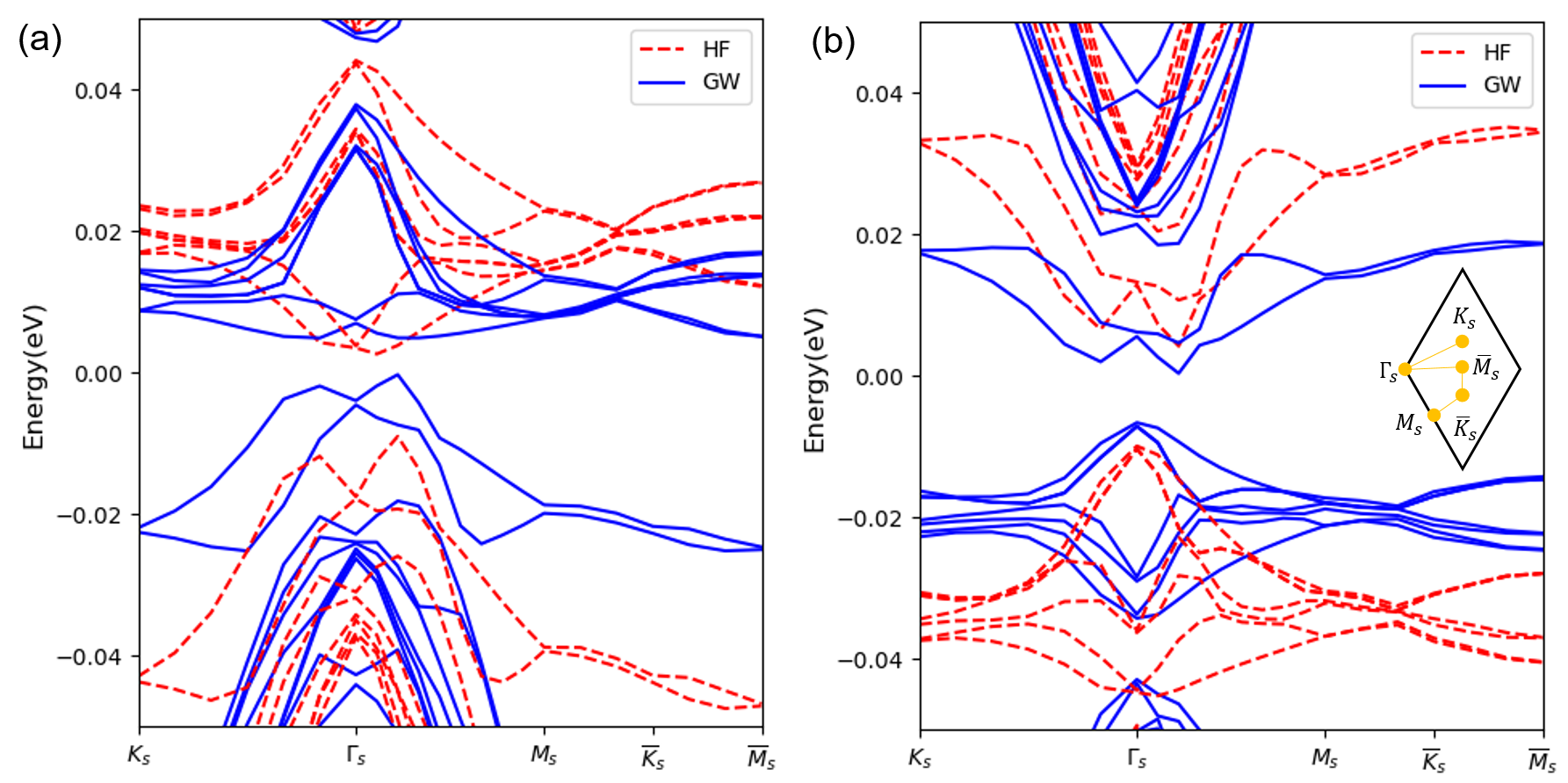}
    \caption{HF and $GW$ band structures of the ground state for TBG of $\theta=1.08^\circ$ at (a) $\nu=-2$ and (b) $\nu= 2$, which turns out to be K-IVC state with Chern number zero. The inset shows the high-symmetry path along which the bands are plotted.}
    \label{fig:TBG_nu2}
\end{figure}

\begin{figure}
    \centering
    \includegraphics[width=0.5\textwidth]{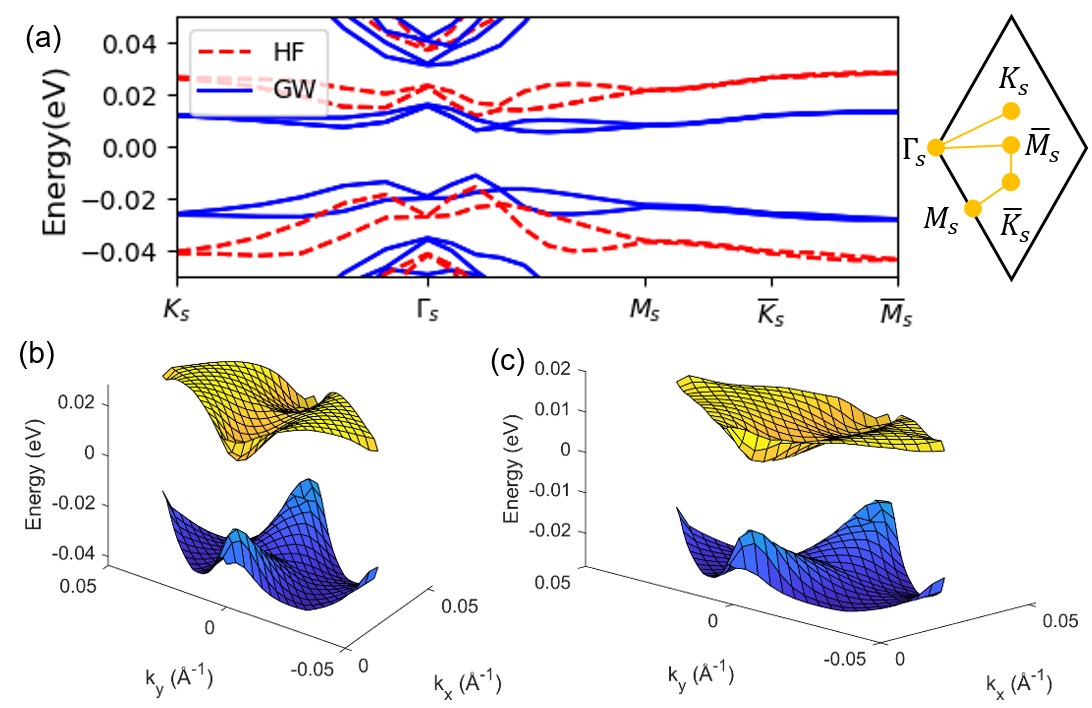}
    \caption{(a) HF and $GW$ band structures of the gapped metastable state for TBG of $\theta=1.08^\circ$ at $\nu=0$, which turns out to be K-IVC state with Chern number zero. The high-symmetry path along which the bands are plotted is given on the right side. (b) and (c) show the energy profile of HF and $GW$ bands in the fist Brillouin zone, respectively.}
    \label{fig:TBG_nu0MSgap}
\end{figure}

\begin{figure}
    \centering
    \includegraphics[width=0.5\textwidth]{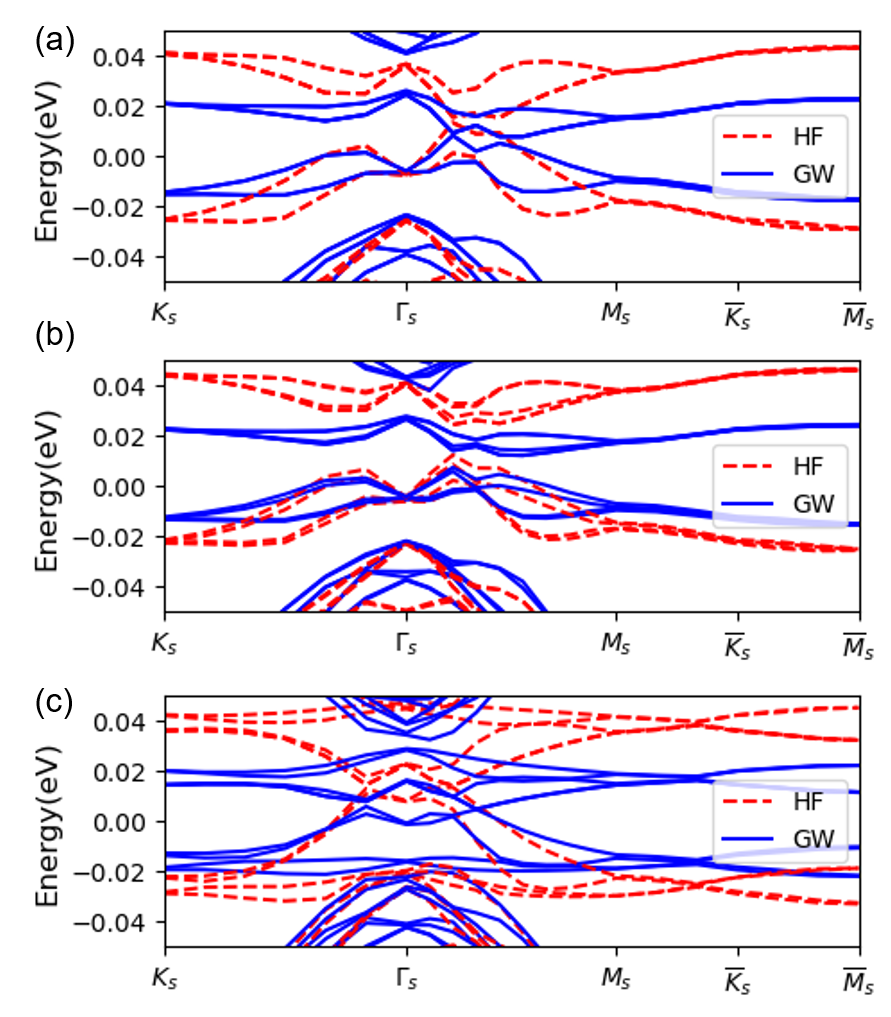}
    \caption{HF and $GW$ band structures of TBG of $\theta=1.08^\circ$ at $\nu=-0.2$: (a) the metallic ground state, obtained by hole doping the nematic metal at $\nu=0$; (b) the metastable state gapped at CNP, which can be seen as a metal obtained by hole doping the $\nu=0$ K-IVC state; (c) the other metallic metastable state, obtained by hole doping the ``symmetric'' metallic metastable state at $\nu=0$.}
    \label{fig:TBG_nu-0.2}
\end{figure}

\bibliography{references}

\end{document}